\documentclass[preprint]{revtex4-1}
\usepackage{graphicx}

\usepackage{natbib}

\usepackage{graphics}

\usepackage{natbib}

\newcommand{\be}{\begin{equation}}
\newcommand{\ee}{\end{equation}}
\newcommand{\nn}{\mbox{} \nonumber \\ \mbox{} }
\newcommand{\ba}{\begin{eqnarray}}
\newcommand{\ea}{\end{eqnarray}}

\newcommand{\E}{{\bf E}}
\newcommand{\B}{{\bf B}}

\newcommand\eg{\textit{e.g.\ }}
\newcommand\cf{\textit{cf.\ }}

\newcommand{\Bf}{{magnetic field}}
\newcommand{\Bfs}{{magnetic fields}}

\newcommand{\Efs}{{electric fields}}

\newcommand{\NS}{neutron star}
\newcommand{\NSs}{{neutron stars}}
\newcommand{\EM}{electromagnetic}
\newcommand{\BH}{{black hole}}
\newcommand{\BHs}{{black holes}}
\newcommand{\Sc}{Schwarzschild}
\newcommand{\ms}{magnetosphere}
\newcommand{\Fermi}{{\it Fermi}}

\begin{document}

\title{Electromagnetic extraction of energy from merging black holes}

\author{Maxim Lyutikov\\
Department of Physics, Purdue University, \\
 525 Northwestern Avenue,
West Lafayette, IN
47907-2036 }

\begin{abstract}
We calculate the evolution of the prompt  intrinsic Poynting power generated by merging black holes.
Orbiting black holes induce rotation of the space-time. In a presence of  magnetic field   supported  by an accretion disk outside of the orbit, this  results in a   generation of an electromagnetic outflow via the Blandford-Znajek-type process with luminosity $L_{EM} \sim   G^3 M^3  B^2/(c^5 R_{orb})  $  and   reaching  a fairly low maximum values  of $L_{EM} = 10^{37}-10^{39} m_6 $ erg s$^{-1}$ ($m_6$ is the masses of black holes in millions of Solar mass) at the time of the  merger.
 Dissipation of the wind power may produce two types of observed signatures: a  highly  variable  collimated emission coming from the internal dissipation within the jets and  a broad-band near-isotropic emission generated
at the termination  shocks.
 \end{abstract}

\maketitle

PACS numbers: 04.30.Tv,  95.85.Sz

\section{Introduction}

Direct detection of gravitational waves of the associated  merging events  of compact objects is one of the prime goals of modern astrophysics. Observations of the corresponding \EM\ signal is most desirable, as it will provide  crucial  information on the location and the physical properties of the event. In case of the merger of stellar-mass compact objects 
(\NSs\ and/or \BHs), the short Gamma Ray Bursts may be the  corresponding  electromagnetic event \cite{1993Natur.361..236M}. 
On the other hand, the detection of the  merger of supermassive black holes in the centers of galaxies  by the  Laser Interferometer Space Antenna,   is an exciting possibility \cite{1980Natur.287..307B,2008ApJ...677.1184L}.  There are clear observational evidence for recent galaxy  mergers \cite{2003ApJ...582..559V}  and direct observations evidence  of binary \BHs\  \cite{2009MNRAS.398L..73D}. 

So far, the studies of a possible prompt  electromagnetic signatures of the \BH\ mergers were attributed mostly  to the perturbations that  the \BHs\ induce in the surrounding gas,  \cite{2002ApJ...567L...9A,2010ApJ...709..774K,2005ApJ...622L..93M,2010ApJ...711L..89V}. The resulting \EM\ signal is  then subject to great uncertainty and   naturally  depends on the complicated non-linear fluid behavior of the system.  One of the problems  is that there should be little gas inside the orbit of the merging  \BHs\  since the  timescale for shrinkage of the binary orbit by gravitational wave radiation becomes shorter than the timescale for mass inflow due to viscose stresses in the disk \cite{2005ApJ...622L..93M}. It is then hard to excite transient dissipative processes in the faraway accretion disk.  In addition, the resulting  long times scales of emission correspond more to the afterglow phase than to  the prompt emission.

 In this paper we take an alternative approach,  which follows  in spirit the seminal paper of   Blandford \& Znajek
 \cite{Blandford:1977}.    Blandford \& Znajek demonstrated that the rotation  of the space-time itself can generate a powerful \EM\ outflow
 carried by a strongly magnetized wind.  In case of merging \BHs\   the rotation  of the space-time is induced  by the orbital motion. 
  The properties of the resulting Poynting flux depend only on the \BHs\ masses, their separation and the assumed \Bf, which in turn can be estimated using a standard anzats of accretion physics. 
 
 Previously,  in  Ref. \cite{2010Sci...329..927P,2010PhRvD..82d4045P} numerical simulations  of the electromagnetic signal during the last orbit before the merger were performed, showing a  production   of collimated Poynting flux  jets. The jets are produced due to linear motion of the \BHs\ though \Bf\ via a mechanism that roughly resembles unipolar induction. The classical analogy of this mechanism is the motion of planet  Io in Jupiter's \Bf\  \cite{1969ApJ...156...59G}.  The mechanism we discuss in the current paper is different, relying on the principles of a Faraday disk, where the rotation of space-time plays a role of the conducing surface, drags the \Bf\ lines and generates \EM\ outflows  \cite{Blandford:1977}.

\section{Faraday  disk in astrophysics}
\label{1}

Many astrophysical sources are powered by the  rotational energy of a  central source using \Bf\ as a "conveyor" belt to transport energy to large distances. Pulsars \cite{GoldreichJulian} are the prime examples of  such process. 
Qualitatively, the  \NS\  surfaces permeated by the \Bf\ works as a Faraday  disk or unipolar inductor, generating a Poynting flux  propagating away from the star. A qualitatively new application of this idea was proposed by Blandford and Znajek \cite{Blandford:1977}, who showed that, formally, there is no need for a hard conducting disk to launch the Poynting flux: rotation of the space itself can effectively act as a Faraday  disk.
In this paper we apply the  ideas of Blandford and Znajek \cite{Blandford:1977} to the space-time produced by orbiting \BHs. 

An important difference between the astrophysical applications mentioned above and the classical Faraday experiment is that in generating the Poynting flux large \Efs\ develop, which  even in the absence of any pre-existing plasma will lead to vacuum breakdown due to various radiative effects. Thus, the system forms the plasma out of the vacuum, so that in general the problem should be describe by relativistic magnetohydrodynamics. 

The magnetic energy density in the low density regions (from where the outflows are generated)  is expected to greatly exceeds the plasma energy density, including rest mass. Charged particles carry current and generate charge density, that ensures ideal MHD condition $\E \cdot \B=0$. In the limit of negligible inertial contribution,  the system reaches a so-called force-free state, where electromagnetic forces balance itself. We stress that this   is not vacuum: inertia-less charge particles ensure ideal MHD conditions. (Formally, in such plasma the stress tensor is diagonalizable; in addition, a condition $|B|>|E|$ must be satisfied.)

\section{\EM\ power of orbiting masses in external \Bf}

Let us consider two black holes of equal mass  $M$ moving  on a circular trajectory of radius $R$. Let us assume that there is a magnetized disk   outside of the orbit that supports the electric currents, which produce a typical \Bf\ $B$ within the orbit. 
We are  interested in the power produced specifically due to the orbital motion of \BHs. 
If {\BH}s are spinning,  each will  in addition act as a Faraday  disk producing the Blandford-Znajek Poynting power  \cite{Blandford:1977}. In addition matter accreting onto \BHs\ will produce some  accretion luminosity both through viscous dissipation in the disk and by launching magnetized winds via the Blandford-Payne-type  \cite{Blandford:1982} mechanism (as  seen in simulations  \cite{2010Sci...329..927P}). 
 Let us give qualitative estimates of the \EM\   power specifically due to the orbital motion of {\BH}s. (Below, for clarity, we omit numerical factors of the order of  a few).

The power produced by the Faraday  disk can be estimated as  
\be
L _{EM}\sim  R^2 B^2 c (\Omega r/c)^2
\ee
where $\Omega$ is the angular velocity of disk rotation. 
In the  case of orbiting \BHs, $\Omega$ is  the typical angular velocity of the rotation of the space-time within the \BH\ orbit. In the weak field regime  the angular velocity of the rotation of space-time  within the orbit can be estimated as \citep[\eg][Eq. 105.20]{LLII}
\be
\Omega \sim {G |{\bf M}| \over c^2 R^3} = { (GM)^{3/2} \over c^2 R^{5/2}}
\label{Omega}
\ee
where $|{\bf M}|  \sim \sqrt{G r} M^{3/2} $ is the total angular momentum of the orbiting {\BH}s (\cf\,  Ref. \cite{Lightman}, problem 13.18). Note, that since outside of the orbit $\Omega \propto R^{-3}$, the  corresponding \EM\  power is small.

\begin{figure}[h!]
   \includegraphics[width=0.95\linewidth]{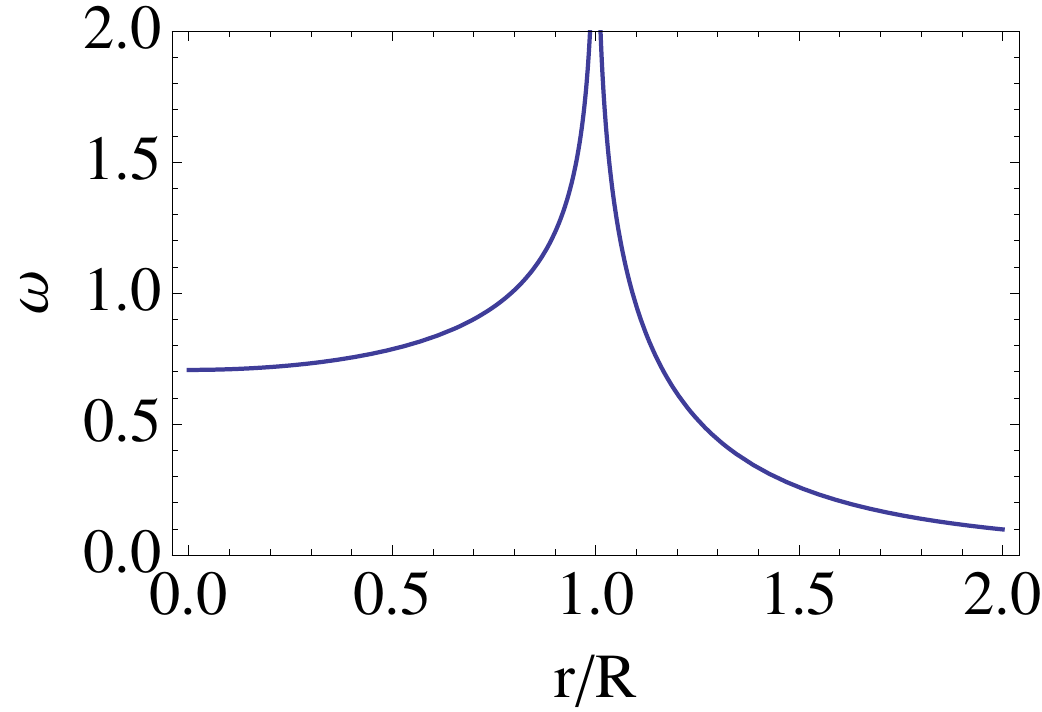}
   \caption{Angular frequency of the rotation of space as a function of radius for a massive ring rotating with Keplerian velocity at radius $R$. Frequency is normalized to 
   $ (GM)^{3/2} / ( c^2 R^{5/2})$. Inside the ring $\Omega$ is nearly constant, falling off $\propto r^{-3}$ outside of the ring. Divergence at $r=R$ is an artifact of the weak field approximation used to calculate $\Omega$.
   }
 \label{omega}
 \end{figure}

Alternatively, we can calculate  the rotation frequency of space-time due to orbiting \BHs\ of equal masses $M$ on a circular orbit if radius $R$ well before the merger, when the motion can be treated in a linearized theory. As a time-averaged  approximation, assume that two \BHs\ may be represented as a ring of mass $2M$ rotating with velocity 
$v=\sqrt{ G M /(2 R)}$, producing a mass current $I_m = \sqrt{G} M^{3/2} / (\sqrt{2} \pi  R^{3/2})$.  In the linearized theory  and assuming  the time-independent field,  the components of the metric tensor satisfy \cite{MTW}
\be
\Delta  g_{0 \phi}=  -16 \pi T_{0 \phi} 
\ee
The component $g_{0 \phi}$ then can be found from the known expression for \Bf\ produced by a current ring \citep[][p. 181]{1975clel.book.....J}
\ba && 
g_{0 \phi} =-  {2 \sqrt{2} \over \pi c^2} (G M/r)^{3/2} { 1\over \sqrt{ r^2 +R^2 + 2 r  R \sin \theta}} \left( \left( 2 \over k^2 -1 \right) K(k^2) - {2 \over k^2} E(k^2) \right)
\nn &&
k^2 = {4 r R \sin \theta \over r^2 +R^2 + 2 r  R \sin \theta}
\ea
where $K$ and $E$ are elliptic integrals. 
The angular frequency of the rotation of space  is then
\be
\Omega = {h_{0 \phi} \over r \sin \theta}
\label{om}
\ee
(Note that there is no singularity at $r=0$, see Fig. \ref{omega}).

In the orbital plane,  $\Omega$  is a slowly varying function of  $r$ with the value at $r=0$ equal to
$\Omega = (GM)^{3/2} / ( \sqrt{2} c^2 R^{5/2})$, in agreement with our estimate (\ref{Omega}),
see Fig. \ref{omega}

Consider next a Faraday disk of radius $R$ embedded in a highly magnetized plasma  in the  force-free approximation discussed in \S \ref{1} and permeated by a constant \Bf\ aligned with the disk normal. The disk is rotating with angular velocity $\Omega$.   The following fields expressed in  cylindrically collimated $r, \phi,z$ then satisfy 
Maxwell's equation and ideal condition $\E \cdot \B=0$ \cite{2001bhgh.book.....P,2002astro.ph..6076K}:
\be
B_r = 0,\, B_\phi=E_r = -{   \Omega r \over c}  B_0, \, B_z =B_0
\ee
for $r< R$ and $B_z =B_0$, $\E=0$ for $r> R$. 
The Poynting flux is then
\be 
L_{EM} = {1\over 8} B_0^2 R^2 c ( R \Omega /c)^2, 
\ee
where $\Omega$ is the  angular velocity of rotation of space (\ref{Omega}), (and {\it not} the angular velocity of the Keplerian rotation $\Omega_K$). 

We stress that in the model problem considered above  (of a rotating disk in a force-free plasma)  {\it  the rotation of field lines at $r<R$ does not break a force balance in  the cylindrical radial   direction $r$, resulting in a formation of cylindrically collimated outflow.} 
We expected that the resulting jets will have a low baryon contamination and thus will be moving at relativistic speeds before they start interacting with the surrounding medium. 

Thus, the \EM\ power due to orbiting \BHs\ acting as a  Faraday  disk  in the external \Bf\ of strength $B$ can be estimated as
\be
L_{EM} \sim {  G^3 M^3 \over c^5 R} B^2 = 6 \times 10^{38} {\rm erg s^{-1}} \,  b_3 ^2 m_6 ^2  \left({R/R_G} \right)^{-1}
\label{LEM}
\ee
where $ b_3 = B/10^3 \, {\rm G}$,  $m_6= M/10^6 M_\odot$, $R_G = GM /c^2 =1.5 \times 10^{11}  {\rm cm} \, m_6$. (In the numerical estimate in  Eq. (\ref{LEM}) and below  the ratio
$\left({r/r_G} \right)$ is treated as a parameter with no dependence on mass.)

(Note, that qualitatively, the power of the Blandford-Znajek process can be similarly estimated using $\Omega = a c/r_G$, 
\be 
L_{BZ} \sim a^2 B^2 r_G^2 c
\label{LBZ}
\ee
where $a$ is the \BH\  spin parameters and $r_G= GM/c^2$ is the \Sc\ radius. At the merger moment $r \sim G M /c^2$, so that Eq. (\ref{LEM}) agrees with Eq. (\ref{LBZ}) for $a \sim 1$, a critically rotating \BH.)

Let us compare the  \EM\ power  to the power emitted in gravitational waves (GWs).
The GW power \cite[][\S 110]{LLII}
\be
L_{GW} \approx {G M^2 \Omega_K^6 R^4 \over c^5} = { G^4 M^5 \over c^5 R^5}
\ee
The ratio of EM to GW luminosities is 
\ba &&
{L _{EM} \over L_{GW}} = {B^2 R^4 \over G M^2} = { E_B \over E_G}
\\ && 
E_B = B^2 R^3
\\ &&
E_G= G M^2/R
\ea
where $E_B$ is the energy of \Bf\ within the orbit of BHs and  $E_G$ is the gravitational potential energy.

The two powers are equal when
\be
r_{\rm eq} \approx {G^{1/4} \sqrt{M} 
\over \sqrt{B}} = 2 \times 10^{16}\, {\rm cm}  \,  b_3^{-1/2} m_6^{1/2}
\label{R}
\ee
For smaller separations, the GW power dominates.
The radius (\ref{R})  is  very large, with a merger time $\sim c^5 r_{\rm eq}^4/(G^3 M^3)$  larger than    the age of the Universe. 

\section{Estimates of \Bf}

In the above estimates we used the parametrization of the \Bf\ somewhat arbitrarily.  In this section we estimate the  \Bf\ within the orbit produced by  the currents flowing  in the accretion disk.  As mentioned in the Introduction, two stages of disk dynamics may be identified. At early times the loss of energy via  gravitational radiation is slow enough, so that due to  viscous diffusion  the inner edge of the disk will be located close to the orbital radius, $R_{\rm d} \approx 2 R$. At later times, for $R_d \leq \xi_d G M /c^2, \, \xi _d \approx40- 100$  \cite{2005ApJ...622L..93M,2009ApJ...700..859O}, the binary will decouple from the disk, undergoing a   merger,  while the inner edge of the disk remains fixed at $R_d$. The decoupling occurs at time
$t_d \sim \xi_d ^4\, G M/c^2 \approx 10 \, {\rm yrs} \,m_6^{-3}$ before the merger.

Several different scalings can be used to estimate the  \Bf\ in the disk and within the orbit. Magnetic field is generated in the disk by the MRI dynamo \cite{BalbusHawley}, which in a thin accretion disk produces fields at equipartition with the turbulent motion, $B^2 \sim \rho v_{turb}^2$. In the standard $\alpha$ disk the  turbulent velocity  is a fraction of the local sound speed $c_s$, $ v_{turb} \sim \sqrt{\alpha } c_s$, while the sound speed $c_s$ is a function of the parameters of the central object, mass flow $\dot{M}$ and location in the disk   \cite{ShakuraSunyaev} (\eg\ in the radiation or matter-dominated parts of the disk).  It is somewhat beyond the scope of this paper to discuss all the possible combinations of the parameters of the accretion disk. Instead, we estimate the upper limits on \Bf\ given various parametrizations of the accretion flow. As we will see, even the upper limits on the \Bfs\ still produce fairly low luminosities, Eq.  (\ref{LLL}).

As a first approximation, we assume that the turbulent velocity is  of the order of the sound speed,  $v_{turb} \sim c_s$; in this case the  \Bf\  is in the     equipartition with the particle energy-densities, $B^2 \approx \rho c_{s}^2$. Secondly, we assume thick accretion disk, so that  the sound speed $c_{s}$  at radius $R$ is of the order of the free fall velocity, $c_{s} \approx \sqrt{ G M/R}$. Two possible  estimates  then may be given for the plasma density $\rho$.
The density can be found either from Bondi scaling,
$\rho R^2 \sqrt{ G M/R} \approx \dot{M} \rho_{ex} (G M )^2 /c_{s,ex}^3$, where $\rho_{ex} $ and $c_{s,ex}$ are density and sound speed in the surrounding medium. This gives
\be
B_{B}= { ( G M )^{5/4} \sqrt{\rho_{ex} } \over R^{5/4} c_{s,ex}^{3/2}} \approx 6 \times  10^4 \, {\rm G}\, n_{-1}^{1/2} c_{s,ex, 6} ^{-3/2} \left( {R \over R_G} \right)^{-5/4}
\ee
where the subscript indicates Bondi scaling and the external number  density is scaled as 
$\rho_{ex} \approx m_p n = m_p (n/0.1 {\rm cm}^{-3}) n_{-1}$ and the external sound speed as 
$c_{s,ex} \approx 10^7 {\rm cm\, s}^{-1} c_{s,ex, 7}$.

Alternatively, the density can be found from the requirement that the accreting matter powers an Eddington-type outflow  $ \eta_M \dot{M} c^2 = L_{\rm Edd}$ with some efficiency $\eta_M\approx 0.1 (\eta_M/0.1) \eta_{M,-1}$:
\be
B_M= { (GM)^{3/4} \sqrt{m_p} \over  \sqrt{\eta_M c \sigma_T} R^{5/4} } \approx  3 \times 10^5 \, {\rm G}\, m_6 ^{-1/2} \eta_{M,-1}^{-1/2} \,  \left( {R \over R_G} \right)^{-5/4}
\ee

Finally, \Bf\ can be estimated assuming that   a fraction of the Eddington luminosity is  carried by \Bf,   $B^2 \approx \eta_E L_{\rm Edd}/( c R^2)$,
\be
B_E =  \eta_{E}^{1/2} { (GM)^{1/2}  \sqrt{m_p} \over   \sqrt{ \sigma_T} R }  \approx    3 \times 10^4 \, {\rm G}\, m_6 ^{-1/2} \eta_{E,-1}^{1/2} \,  \left( {R \over R_G} \right)^{-1}
\ee

In all cases the \Bf\ within the orbit  should be evaluated at the inner edge of the accretion disk, located either at $R_d \sim R$ before decoupling or at 
$R_d \approx 100 R_G$ after decoupling. Accordingly, we find
\ba &&
L_B = \left\{ 
\begin{array}{ll}
 { ( G M)^{11/2} \rho_{\rm ex} \over c^5 c_{s,ex}^3 R^{7/2} }&\mbox{for  }  {R \over G M/c^2} > \xi_d \\
  {  ( G M)^{3} \rho_{\rm ex} \over  c_{s,ex}^3 R  \xi_d^{5/2} }& \mbox{for  } {R \over G M/c^2} <  \xi_d 
  \end{array} \right.\, for B =B_B
  \nn &&
L _M= \left\{ 
\begin{array}{ll}
 { ( G M)^{9/2} m_p \over c^6 \eta_M R^{7/2} \sigma_T} &\mbox{for  }  {R \over G M/c^2} > \xi_d \\
 {  ( G M)^2 c m_p \over  \eta_M \xi_d^{5/2}  c R \sigma_T} & \mbox{for  } {R \over G M/c^2} <  \xi_d 
  \end{array} \right. \, for B =B_M
 \nn &&
L_E = \left\{ 
\begin{array}{ll}
\eta_E  { ( G M)^{4} m_p  \over c^5  R^{3} \sigma_T} &\mbox{for  }  {R \over G M/c^2} > \xi_d \\
\eta_E {   (G M)^2  m_p \over \xi_d^{2} \sigma_T cR} & \mbox{for  } {R \over G M/c^2} <  \xi_d 
  \end{array} \right. \,  for B =B_E
  \label{Lmax}
  \ea
 The maximal powers, reached at the time of the merger are
 \be
 L_B \sim 10^{37} \, {\rm erg s}^{-1} m_6^2 n_{-1} c_{s,ex,6}^{-3} , \, 
 L_M \sim 10^{39}   \, {\rm erg s}^{-1} m_6 \eta_{M,-1}^{-1} , \, 
 L_E \sim 10^{38}  \, {\rm erg s}^{-1}  m_6 \eta_{E,-1}
 \label{LLL}
 \ee
 Thus, various parametrizations of the \Bf\ give a fairly consistent value of the \EM\ luminocity.
 
  The temporal evolution of  the luminosity in each case can be found from the equation for the 
 orbital separation at time $-t$ before the merger  \cite[][\S 110]{LLII},
\be
R= (  (G M)^3 (-t)/c^5)^{1/4}
\ee
 The total energies emitted after the decoupling are 
 \ba &&
 E_B \approx { (G M)^3 \rho_{\rm ex} \over c c_s^3} \approx 10^{43} {\rm erg}
 \nn &&
 E_M \approx  { (G M)^2 m_p \sqrt{\xi_d} \over \eta_M c^2 \sigma_T}  \approx 5 \times 10^{44} {\rm erg}
 \nn &&
 E_E \approx   { (G M)^2 m_p \xi_d \eta_E  \over  c^2 \sigma_T}  \approx 5 \times 10^{42} {\rm erg}
 \label{Lt}
 \ea

\section{Expected emission}

We foresee two types of electromagnetic signals coming from merging of  \BHs. First, internal dissipation within the jets can produce highly boosted emission similar to blazar jets and high energy emission from Gamma Ray Bursts (GRBs). (Blazars are types of Active  Galactic Nuclei  (AGNs) with relativistic  jets directed nearly  straight at the  observer.) Blazars dominate the  extragalactic sources in the \Fermi\ catalogue \cite{2010ApJ...722..520A,2010A&A...512A..24S}. This type of emission will be highly variable due to relativistic boosting, similar to prompt GRB emission \cite{FermiGRB080916C}. 
On the other hand, internal jet   emission will be highly  anisotropic, mostly aberrated   along the direction of motion of the jets - perpendicular to the binary orbit. Thus, this type of emission will be detected only if the line of sight to the binary is nearly aligned with the binary normal.  

Secondly, when the jet starts to  interact with the circumbinary medium, its kinetic power will be dissipated in a termination shock, similar to the case of pulsar wind nebulae (magnetically-dominated jets, which lack strong termination shock, may produce similar observational signatures \cite{2010MNRAS.405.1809L}.)  The post-shock flow will be  only  weakly relativistic, producing nearly isotropic emission. 

By analogy with the emission from PWNe, one expects that a broad range of frequencies is produced by the Poynting-flux dominated wind. Unfortunately, our understanding of the dissipation and accelerations processes in such systems is not sufficient to make quantitative predictions. By comparison, \eg\ with the Crab nebular, we  expect that that the peak of the spectral energy density falls into the optical-UV-soft X-ray band, with the total dissipated power reaching tens of percent of the Poynting power.

\section{Discussion}

We have calculated the \EM\ power expected from the merging \BHs. The expected Poynting scales {\it not} as a  square of the orbital frequency, but as a square of the rotational frequency of the space-time, Eq. (\ref{Omega}). The maximum power is  reached during the final merger orbit, when the orbital frequency and the rotational frequency of the space-time are nearly equal, since the resulting \BH\ has the spin parameter of the order of unity. Still, the distinction between the two frequencies is a principal point in our approach.

 Overall, the corresponding peak  Poynting powers  (\ref{LLL}) are fairly low, even though the total emitted energies are reasonably high (\ref{Lt}). In addition, it is expected that only a fraction of the Poynting power is converted into radiation.   Still, there is a chance of detection  in the 
 fortuitous case when the line of sight to the system is nearly aligned with the orbital normal and if the outflow reaches relativistic velocities. In this case  most of the \EM\ power will beamed along the direction of motion.

I would like to thank Scott Hughes and Milos Milosavljevic. 

\bibliographystyle{apsrev}
\bibliography{/Users/maxim/Home/Research/BibTex}

\begin{thebibliography}{28}
\expandafter\ifx\csname natexlab\endcsname\relax\def\natexlab#1{#1}\fi
\expandafter\ifx\csname bibnamefont\endcsname\relax
  \def\bibnamefont#1{#1}\fi
\expandafter\ifx\csname bibfnamefont\endcsname\relax
  \def\bibfnamefont#1{#1}\fi
\expandafter\ifx\csname citenamefont\endcsname\relax
  \def\citenamefont#1{#1}\fi
\expandafter\ifx\csname url\endcsname\relax
  \def\url#1{\texttt{#1}}\fi
\expandafter\ifx\csname urlprefix\endcsname\relax\def\urlprefix{URL }\fi
\providecommand{\bibinfo}[2]{#2}
\providecommand{\eprint}[2][]{\url{#2}}

\bibitem[{\citenamefont{{Mochkovitch} et~al.}(1993)\citenamefont{{Mochkovitch},
  {Hernanz}, {Isern}, and {Martin}}}]{1993Natur.361..236M}
\bibinfo{author}{\bibfnamefont{R.}~\bibnamefont{{Mochkovitch}}},
  \bibinfo{author}{\bibfnamefont{M.}~\bibnamefont{{Hernanz}}},
  \bibinfo{author}{\bibfnamefont{J.}~\bibnamefont{{Isern}}}, \bibnamefont{and}
  \bibinfo{author}{\bibfnamefont{X.}~\bibnamefont{{Martin}}},
  \bibinfo{journal}{\nat} \textbf{\bibinfo{volume}{361}}, \bibinfo{pages}{236}
  (\bibinfo{year}{1993}).

\bibitem[{\citenamefont{{Begelman} et~al.}(1980)\citenamefont{{Begelman},
  {Blandford}, and {Rees}}}]{1980Natur.287..307B}
\bibinfo{author}{\bibfnamefont{M.~C.} \bibnamefont{{Begelman}}},
  \bibinfo{author}{\bibfnamefont{R.~D.} \bibnamefont{{Blandford}}},
  \bibnamefont{and} \bibinfo{author}{\bibfnamefont{M.~J.}
  \bibnamefont{{Rees}}}, \bibinfo{journal}{\nat}
  \textbf{\bibinfo{volume}{287}}, \bibinfo{pages}{307} (\bibinfo{year}{1980}).

\bibitem[{\citenamefont{{Lang} and {Hughes}}(2008)}]{2008ApJ...677.1184L}
\bibinfo{author}{\bibfnamefont{R.~N.} \bibnamefont{{Lang}}} \bibnamefont{and}
  \bibinfo{author}{\bibfnamefont{S.~A.} \bibnamefont{{Hughes}}},
  \bibinfo{journal}{\apj} \textbf{\bibinfo{volume}{677}}, \bibinfo{pages}{1184}
  (\bibinfo{year}{2008}), \eprint{0710.3795}.

\bibitem[{\citenamefont{{Volonteri} et~al.}(2003)\citenamefont{{Volonteri},
  {Haardt}, and {Madau}}}]{2003ApJ...582..559V}
\bibinfo{author}{\bibfnamefont{M.}~\bibnamefont{{Volonteri}}},
  \bibinfo{author}{\bibfnamefont{F.}~\bibnamefont{{Haardt}}}, \bibnamefont{and}
  \bibinfo{author}{\bibfnamefont{P.}~\bibnamefont{{Madau}}},
  \bibinfo{journal}{\apj} \textbf{\bibinfo{volume}{582}}, \bibinfo{pages}{559}
  (\bibinfo{year}{2003}), \eprint{arXiv:astro-ph/0207276}.

\bibitem[{\citenamefont{{Dotti} et~al.}(2009)\citenamefont{{Dotti}, {Montuori},
  {Decarli}, {Volonteri}, {Colpi}, and {Haardt}}}]{2009MNRAS.398L..73D}
\bibinfo{author}{\bibfnamefont{M.}~\bibnamefont{{Dotti}}},
  \bibinfo{author}{\bibfnamefont{C.}~\bibnamefont{{Montuori}}},
  \bibinfo{author}{\bibfnamefont{R.}~\bibnamefont{{Decarli}}},
  \bibinfo{author}{\bibfnamefont{M.}~\bibnamefont{{Volonteri}}},
  \bibinfo{author}{\bibfnamefont{M.}~\bibnamefont{{Colpi}}}, \bibnamefont{and}
  \bibinfo{author}{\bibfnamefont{F.}~\bibnamefont{{Haardt}}},
  \bibinfo{journal}{\mnras} \textbf{\bibinfo{volume}{398}},
  \bibinfo{pages}{L73} (\bibinfo{year}{2009}), \eprint{0809.3446}.

\bibitem[{\citenamefont{{Armitage} and
  {Natarajan}}(2002)}]{2002ApJ...567L...9A}
\bibinfo{author}{\bibfnamefont{P.~J.} \bibnamefont{{Armitage}}}
  \bibnamefont{and}
  \bibinfo{author}{\bibfnamefont{P.}~\bibnamefont{{Natarajan}}},
  \bibinfo{journal}{\apjl} \textbf{\bibinfo{volume}{567}}, \bibinfo{pages}{L9}
  (\bibinfo{year}{2002}), \eprint{arXiv:astro-ph/0201318}.

\bibitem[{\citenamefont{{Krolik}}(2010)}]{2010ApJ...709..774K}
\bibinfo{author}{\bibfnamefont{J.~H.} \bibnamefont{{Krolik}}},
  \bibinfo{journal}{\apj} \textbf{\bibinfo{volume}{709}}, \bibinfo{pages}{774}
  (\bibinfo{year}{2010}), \eprint{0911.5711}.

\bibitem[{\citenamefont{{Milosavljevi{\'c}} and
  {Phinney}}(2005)}]{2005ApJ...622L..93M}
\bibinfo{author}{\bibfnamefont{M.}~\bibnamefont{{Milosavljevi{\'c}}}}
  \bibnamefont{and} \bibinfo{author}{\bibfnamefont{E.~S.}
  \bibnamefont{{Phinney}}}, \bibinfo{journal}{\apjl}
  \textbf{\bibinfo{volume}{622}}, \bibinfo{pages}{L93} (\bibinfo{year}{2005}),
  \eprint{arXiv:astro-ph/0410343}.

\bibitem[{\citenamefont{{van Meter} et~al.}(2010)\citenamefont{{van Meter},
  {Wise}, {Miller}, {Reynolds}, {Centrella}, {Baker}, {Boggs}, {Kelly}, and
  {McWilliams}}}]{2010ApJ...711L..89V}
\bibinfo{author}{\bibfnamefont{J.~R.} \bibnamefont{{van Meter}}},
  \bibinfo{author}{\bibfnamefont{J.~H.} \bibnamefont{{Wise}}},
  \bibinfo{author}{\bibfnamefont{M.~C.} \bibnamefont{{Miller}}},
  \bibinfo{author}{\bibfnamefont{C.~S.} \bibnamefont{{Reynolds}}},
  \bibinfo{author}{\bibfnamefont{J.}~\bibnamefont{{Centrella}}},
  \bibinfo{author}{\bibfnamefont{J.~G.} \bibnamefont{{Baker}}},
  \bibinfo{author}{\bibfnamefont{W.~D.} \bibnamefont{{Boggs}}},
  \bibinfo{author}{\bibfnamefont{B.~J.} \bibnamefont{{Kelly}}},
  \bibnamefont{and} \bibinfo{author}{\bibfnamefont{S.~T.}
  \bibnamefont{{McWilliams}}}, \bibinfo{journal}{\apjl}
  \textbf{\bibinfo{volume}{711}}, \bibinfo{pages}{L89} (\bibinfo{year}{2010}),
  \eprint{0908.0023}.

\bibitem[{\citenamefont{{Blandford} and {Znajek}}(1977)}]{Blandford:1977}
\bibinfo{author}{\bibfnamefont{R.~D.} \bibnamefont{{Blandford}}}
  \bibnamefont{and} \bibinfo{author}{\bibfnamefont{R.~L.}
  \bibnamefont{{Znajek}}}, \bibinfo{journal}{MNRAS}
  \textbf{\bibinfo{volume}{179}}, \bibinfo{pages}{433} (\bibinfo{year}{1977}).

\bibitem[{\citenamefont{{Palenzuela}
  et~al.}(2010{\natexlab{a}})\citenamefont{{Palenzuela}, {Lehner}, and
  {Liebling}}}]{2010Sci...329..927P}
\bibinfo{author}{\bibfnamefont{C.}~\bibnamefont{{Palenzuela}}},
  \bibinfo{author}{\bibfnamefont{L.}~\bibnamefont{{Lehner}}}, \bibnamefont{and}
  \bibinfo{author}{\bibfnamefont{S.~L.} \bibnamefont{{Liebling}}},
  \bibinfo{journal}{Science} \textbf{\bibinfo{volume}{329}},
  \bibinfo{pages}{927} (\bibinfo{year}{2010}{\natexlab{a}}),
  \eprint{1005.1067}.

\bibitem[{\citenamefont{{Palenzuela}
  et~al.}(2010{\natexlab{b}})\citenamefont{{Palenzuela}, {Garrett}, {Lehner},
  and {Liebling}}}]{2010PhRvD..82d4045P}
\bibinfo{author}{\bibfnamefont{C.}~\bibnamefont{{Palenzuela}}},
  \bibinfo{author}{\bibfnamefont{T.}~\bibnamefont{{Garrett}}},
  \bibinfo{author}{\bibfnamefont{L.}~\bibnamefont{{Lehner}}}, \bibnamefont{and}
  \bibinfo{author}{\bibfnamefont{S.~L.} \bibnamefont{{Liebling}}},
  \bibinfo{journal}{\prd} \textbf{\bibinfo{volume}{82}},
  \bibinfo{pages}{044045} (\bibinfo{year}{2010}{\natexlab{b}}),
  \eprint{1007.1198}.

\bibitem[{\citenamefont{{Goldreich} and
  {Lynden-Bell}}(1969)}]{1969ApJ...156...59G}
\bibinfo{author}{\bibfnamefont{P.}~\bibnamefont{{Goldreich}}} \bibnamefont{and}
  \bibinfo{author}{\bibfnamefont{D.}~\bibnamefont{{Lynden-Bell}}},
  \bibinfo{journal}{\apj} \textbf{\bibinfo{volume}{156}}, \bibinfo{pages}{59}
  (\bibinfo{year}{1969}).

\bibitem[{\citenamefont{{Goldreich} and {Julian}}(1969)}]{GoldreichJulian}
\bibinfo{author}{\bibfnamefont{P.}~\bibnamefont{{Goldreich}}} \bibnamefont{and}
  \bibinfo{author}{\bibfnamefont{W.~H.} \bibnamefont{{Julian}}},
  \bibinfo{journal}{\apj} \textbf{\bibinfo{volume}{157}}, \bibinfo{pages}{869}
  (\bibinfo{year}{1969}).

\bibitem[{\citenamefont{Blandford and Payne}(1982)}]{Blandford:1982}
\bibinfo{author}{\bibnamefont{Blandford}} \bibnamefont{and}
  \bibinfo{author}{\bibnamefont{Payne}}, \bibinfo{journal}{MNRAS}
  \textbf{\bibinfo{volume}{199}}, \bibinfo{pages}{883} (\bibinfo{year}{1982}).

\bibitem[{\citenamefont{{Landau} and {Lifshitz}}(1971)}]{LLII}
\bibinfo{author}{\bibfnamefont{L.~D.} \bibnamefont{{Landau}}} \bibnamefont{and}
  \bibinfo{author}{\bibfnamefont{E.~M.} \bibnamefont{{Lifshitz}}},
  \emph{\bibinfo{title}{{The classical theory of fields}}}
  (\bibinfo{year}{1971}).

\bibitem[{\citenamefont{{Lightman} et~al.}(1979)\citenamefont{{Lightman},
  {Press}, {Price}, and {Teukolsky}}}]{Lightman}
\bibinfo{author}{\bibfnamefont{A.~P.} \bibnamefont{{Lightman}}},
  \bibinfo{author}{\bibfnamefont{W.~H.} \bibnamefont{{Press}}},
  \bibinfo{author}{\bibfnamefont{R.~H.} \bibnamefont{{Price}}},
  \bibnamefont{and} \bibinfo{author}{\bibfnamefont{S.~A.}
  \bibnamefont{{Teukolsky}}}, \emph{\bibinfo{title}{{Problem book in relativity
  and gravitation.}}} (\bibinfo{year}{1979}).

\bibitem[{\citenamefont{{Misner} et~al.}(1973)\citenamefont{{Misner}, {Thorne},
  and {Wheeler}}}]{MTW}
\bibinfo{author}{\bibfnamefont{C.~W.} \bibnamefont{{Misner}}},
  \bibinfo{author}{\bibfnamefont{K.~S.} \bibnamefont{{Thorne}}},
  \bibnamefont{and} \bibinfo{author}{\bibfnamefont{J.~A.}
  \bibnamefont{{Wheeler}}}, \emph{\bibinfo{title}{{Gravitation}}}
  (\bibinfo{publisher}{San Francisco: W.H.~Freeman and Co., 1973},
  \bibinfo{year}{1973}).

\bibitem[{\citenamefont{{Jackson}}(1975)}]{1975clel.book.....J}
\bibinfo{author}{\bibfnamefont{J.~D.} \bibnamefont{{Jackson}}},
  \emph{\bibinfo{title}{{Classical electrodynamics}}} (\bibinfo{year}{1975}).

\bibitem[{\citenamefont{{Punsly}}(2001)}]{2001bhgh.book.....P}
\bibinfo{author}{\bibfnamefont{B.}~\bibnamefont{{Punsly}}},
  \emph{\bibinfo{title}{{Black hole gravitohydromagnetics}}}
  (\bibinfo{year}{2001}).

\bibitem[{\citenamefont{{Komissarov}}(2002)}]{2002astro.ph..6076K}
\bibinfo{author}{\bibfnamefont{S.~S.} \bibnamefont{{Komissarov}}},
  \bibinfo{journal}{ArXiv Astrophysics e-prints}  (\bibinfo{year}{2002}),
  \eprint{arXiv:astro-ph/0206076}.

\bibitem[{\citenamefont{{O'Neill} et~al.}(2009)\citenamefont{{O'Neill},
  {Miller}, {Bogdanovi{\'c}}, {Reynolds}, and
  {Schnittman}}}]{2009ApJ...700..859O}
\bibinfo{author}{\bibfnamefont{S.~M.} \bibnamefont{{O'Neill}}},
  \bibinfo{author}{\bibfnamefont{M.~C.} \bibnamefont{{Miller}}},
  \bibinfo{author}{\bibfnamefont{T.}~\bibnamefont{{Bogdanovi{\'c}}}},
  \bibinfo{author}{\bibfnamefont{C.~S.} \bibnamefont{{Reynolds}}},
  \bibnamefont{and} \bibinfo{author}{\bibfnamefont{J.~D.}
  \bibnamefont{{Schnittman}}}, \bibinfo{journal}{\apj}
  \textbf{\bibinfo{volume}{700}}, \bibinfo{pages}{859} (\bibinfo{year}{2009}),
  \eprint{0812.4874}.

\bibitem[{\citenamefont{{Balbus} and {Hawley}}(1991)}]{BalbusHawley}
\bibinfo{author}{\bibfnamefont{S.~A.} \bibnamefont{{Balbus}}} \bibnamefont{and}
  \bibinfo{author}{\bibfnamefont{J.~F.} \bibnamefont{{Hawley}}},
  \bibinfo{journal}{\apj} \textbf{\bibinfo{volume}{376}}, \bibinfo{pages}{214}
  (\bibinfo{year}{1991}).

\bibitem[{\citenamefont{{Shakura} and {Sunyaev}}(1973)}]{ShakuraSunyaev}
\bibinfo{author}{\bibfnamefont{N.~I.} \bibnamefont{{Shakura}}}
  \bibnamefont{and} \bibinfo{author}{\bibfnamefont{R.~A.}
  \bibnamefont{{Sunyaev}}}, \bibinfo{journal}{\aap}
  \textbf{\bibinfo{volume}{24}}, \bibinfo{pages}{337} (\bibinfo{year}{1973}).

\bibitem[{\citenamefont{{Abdo}}(2010)}]{2010ApJ...722..520A}
\bibinfo{author}{\bibfnamefont{A.~A.~{\etal}.} \bibnamefont{{Abdo}}},
  \bibinfo{journal}{\apj} \textbf{\bibinfo{volume}{722}}, \bibinfo{pages}{520}
  (\bibinfo{year}{2010}), \eprint{1004.0348}.

\bibitem[{\citenamefont{{Savolainen} et~al.}(2010)\citenamefont{{Savolainen},
  {Homan}, {Hovatta}, {Kadler}, {Kovalev}, {Lister}, {Ros}, and
  {Zensus}}}]{2010A&A...512A..24S}
\bibinfo{author}{\bibfnamefont{T.}~\bibnamefont{{Savolainen}}},
  \bibinfo{author}{\bibfnamefont{D.~C.} \bibnamefont{{Homan}}},
  \bibinfo{author}{\bibfnamefont{T.}~\bibnamefont{{Hovatta}}},
  \bibinfo{author}{\bibfnamefont{M.}~\bibnamefont{{Kadler}}},
  \bibinfo{author}{\bibfnamefont{Y.~Y.} \bibnamefont{{Kovalev}}},
  \bibinfo{author}{\bibfnamefont{M.~L.} \bibnamefont{{Lister}}},
  \bibinfo{author}{\bibfnamefont{E.}~\bibnamefont{{Ros}}}, \bibnamefont{and}
  \bibinfo{author}{\bibfnamefont{J.~A.} \bibnamefont{{Zensus}}},
  \bibinfo{journal}{\aap} \textbf{\bibinfo{volume}{512}}, \bibinfo{pages}{A24+}
  (\bibinfo{year}{2010}), \eprint{0911.4924}.

\bibitem[{\citenamefont{{Abdo}}(2009)}]{FermiGRB080916C}
\bibinfo{author}{\bibfnamefont{A.~A.~{\it {\etal}}.} \bibnamefont{{Abdo}}},
  \bibinfo{journal}{Science} \textbf{\bibinfo{volume}{323}},
  \bibinfo{pages}{1688} (\bibinfo{year}{2009}).

\bibitem[{\citenamefont{{Lyutikov}}(2010)}]{2010MNRAS.405.1809L}
\bibinfo{author}{\bibfnamefont{M.}~\bibnamefont{{Lyutikov}}},
  \bibinfo{journal}{\mnras} \textbf{\bibinfo{volume}{405}},
  \bibinfo{pages}{1809} (\bibinfo{year}{2010}), \eprint{0911.0324}.

\end{thebibliography}

\end{document}